\documentclass[11pt]{article}
\usepackage{jheppub}
\usepackage{mymacros}

\begin{document}
\title{The centaur-algebra of observables}

\author[a]{Sergio E. Aguilar-Gutierrez,}
\author[b]{Eyoab Bahiru,}
\author[c]{Ricardo Espíndola}

\affiliation[a]{Institute for Theoretical Physics, KU Leuven,\\ Celestijnenlaan 200D, B-3001 Leuven, Belgium}
\affiliation[b]{SISSA, International School for Advanced Studies, via Bonomea 265, 34136 Trieste, Italy\\
INFN, Sezione di Trieste, via Valerio 2, 34127 Trieste, Italy\\
International Centre for Theoretical Physics, Strada Costiera 11, 34151 Trieste Italy}
\affiliation[c]{Institute for Advanced Study, Tsinghua University, 100084 Beijing, China}

\emailAdd{sergio.ernesto.aguilar@gmail.com,ebahiru@sissa.it, ricardo.esro1@gmail.com}

\abstract{This letter explores a transition in the type of von Neumann algebra for asymptotically AdS spacetimes from the implementations of the different gravitational constraints. We denote it as the \emph{centaur-algebra} of observables. In the first part of the letter, we employ a class of flow geometries interpolating between AdS$_2$ and dS$_2$ spaces, the centaur geometries. We study the type II$_\infty$ crossed product algebra describing the semiclassical gravitational theory, and we explore the algebra of bounded sub-regions in the bulk theory following $T\overline{T}$ deformations of the geometry and study the gravitational constraints with respect to the quasi-local Brown-York energy of the system at a finite cutoff. In the second part, we study arbitrary asymptotically AdS spacetimes, where we implement the boundary protocol of an infalling observer modeled as a probe black hole proposed by \cite{deBoer:2022zps} to study modifications in the algebra. In both situations, we show how incorporating the constraints requires a type II$_1$ description.}

\maketitle

\section{Introduction}

Recently, there has been interest in the understanding of perturbative quantum gravity in terms of the algebra of diffeomorphism invariant observables, which have allowed us to rigorously define density matrices and the associated notion of generalized entropies\cite{Witten:2021unn,Chandrasekaran:2022cip,Chandrasekaran:2022eqq,Jensen:2023yxy}. Pioneering work developing bulk emergence from the language of von Neumann algebra can be found in \cite{Papadodimas:2013jku,Jefferson:2018ksk,Leutheusser:2021frk,Witten:2021unn} and check \cite{Witten:2018zxz,Witten:2021jzq,Sorce:2023fdx,Casini:2022rlv} for reviews.

This procedure begins with a type III$_1$ algebra describing the quantum fluctuations on a subregion of curved spacetime background. One incorporates dynamical gravity perturbatively by requiring that time translations (among other symmetry generators) act as gauge redundancies, which we denote throughout the letter as gravitational constraints. In the several examples considered, once gravitational corrections are included (either perturbatively or as an addition of a gravitational mode \cite{Chandrasekaran:2022eqq}), it has been shown that the algebra of observables becomes type II$_\infty$ when the gravitational dressing of operators is performed with respect to the asymptotic boundary region of a spatially open universe; while if the dressing is with respect to a worldline observer in a spatially closed universe, the algebra is of type II$_1$ \cite{Witten:2021unn,Chandrasekaran:2022cip,Chandrasekaran:2022eqq,Jensen:2023yxy}. More recently, the construction has been understood in \cite{AliAhmad:2023etg} to be more general in that it does not necessarily rely on gravitational constraints at all.

Let us denote $\mathcal{A}$ as the algebra of bulk fluctuations associated with a spacetime region, acting on a Hilbert space $\mathcal{H}$, and let $T$ be the generator of the unitary representation of automorphism group of $\mathcal{A}$ (for simplicity we take it to be $\mathbb{R}$) on $\mathcal{H}$, with the respective group elements $\qty{U=\rme^{\rmi sT}, ~ \forall s\in\mathbb{R}}$ acting on the algebra elements as, 
\begin{equation}\label{eq:U ops}
    U\,a\,U^{-1}\in \mathcal{A}\,,\quad\forall a\in \mathcal{A}~.
\end{equation}
Let $X$ be the generator of the unitary representation of the automorphism group (of $\mathcal{A}'$) acting on $L^2(\mathbb{R})$\cite{Chandrasekaran:2022cip,Witten:2021unn,takesaki1973duality}. Then, one denotes the crossed product algebra of $\mathcal{A}$ and its automorphism group, $\mathbb{R}$, as
\begin{equation}\label{eq:crossed prod}
\hat{\mathcal{A}}=\mathcal{A}\rtimes\mathbb{R}~,
\end{equation}
which is `produced' by adjoining bounded functions of $T+X$ to $\mathcal{A}$, i.e. 
\begin{equation}\label{eq:crossed prod elements}
    a\rme^{\rmi s T}\otimes\rme^{\rmi s X}\in \hat{\mathcal{A}},\quad \forall a\in \mathcal{A}~,
\end{equation}
acting on a Hilbert space $\hat{\mathcal{H}}\equiv\mathcal{H}\otimes L^2(\mathbb{R})$. When the automorphism is outer, i.e. $U\notin\mathcal{A}$, and $\mathcal{A}$ is a type III$_1$ algebra, the crossed product algebra results in a type II algebra \cite{takesaki1973duality}. Trace-class elements in type II algebras are defined as those with a well-defined trace. For each state $\ket{\Phi} \in \hat{\mathcal{H}}$, the expectation value of any operator $\hat{a}$ is given by $\bra{\Phi}\hat{a}\ket{\Phi}$. Since the trace is non-degenerate, we can associate a positive operator $\rho_{\Phi}$ to the state $\ket{\Phi}$ which reproduces this expectation value,
\begin{equation}
\Tr(\rho_{\Phi}\hat{a})=\bra{\Phi}\hat{a}\ket{\Phi}.
\end{equation}
The operator $\rho_{\Phi}$ is the density matrix of $\ket{\Phi}$ and, a von Neumann entropy can be defined as,
\begin{equation}
    S = -\Tr(\rho_{\Phi}\text{log}\rho_{\Phi})~.
\end{equation}
For semiclassical states in $\hat{\mathcal{H}}$, this entropy was shown to match with the generalized entropy\footnote{More precisely, what is matched is the entropy differences since the von Neumann entropy is defined up to a state-independent additive constant.}\cite{Chandrasekaran:2022eqq}. In the context of the eternal anti-de Sitter (AdS) black hole, the generator of the automorphism group, $T$, is proportional to the time translation generator on both of the asymptotic boundaries. 

One needs different regularization procedures for $T$ depending on whether the systems are described by a canonical \cite{Witten:2021unn} or micro-canonical ensemble \cite{Chandrasekaran:2022eqq}.
More precisely, one should divide the generator by $N$ in the canonical ensemble compared with the micro-canonical ensemble and do the construction in a perturbative series in $\sqrt{G_N} \sim 1/N$ (where $N$ is the rank of the gauge group of the boundary CFT). The reason is that the states in the canonical ensemble have $O(N)$ variance in the energy which diverges in the large $N$ limit. The operator $T+X$ is taken to be the Hamiltonian of the CFT and thus the crossed product algebra $\mathcal{\hat{A}}$ actually describes the physical theory. These methods have also been developed for subregion algebras \cite{Leutheusser:2022bgi,AliAhmad:2023etg,Klinger:2023tgi,Bahiru:2022mwh}. \footnote{Meanwhile, there are some expectations that non-perturbative corrections in quantum gravity might modify the algebra to type I once string theory corrections and black hole microstates are added in the algebra \cite{Witten:2021unn,Dabholkar:2023ows}.} See \cite{Furuya:2023fei,DiGiulio:2023nvz,Banerjee:2023eew,Ali:2023bmm} for related developments in this area.

Physical observables in perturbative quantum gravity are required to be diffeomorphism invariant. For spatially open universes, this is naturally implemented by dressing the operators with respect to the asymptotic boundary\cite{DeWitt:1962cg,Giddings:2005id,Marolf:2015jha}. The reader is referred to \cite{Bahiru:2023zlc,Bahiru:2022oas} for an alternative dressing of the operators with respect to the features of the state itself. Since in a gravitational theory the Hamiltonian is a boundary quantity, this dressing implies that the operators will not commute with the ADM Hamiltonian in general. On the other hand, for spatially closed universes like de Sitter (dS) space and subregions in an open universe, it was proposed in \cite{Chandrasekaran:2022cip,Jensen:2023yxy} that one should perform the dressing with respect to the world-line of an observer. Thus, the dressed observables will translate under the action of the world-line energy of the observer. Both of these facts are encoded in the non-trivial action of $T+X$ on the elements of $\mathcal{A}$.\footnote{Both in \cite{Chandrasekaran:2022cip} and \cite{Jensen:2023yxy} the dressed observables are not given in the terms of the elements given in \ref{eq:crossed prod elements}, rather in an equivalent description where the elements are $e^{iTP}ae^{-iTP}$ and $e^{isX}$, where $P$ is the conjugate variable to $X$, which is taken to be the energy of the observer. This description can be related to \ref{eq:crossed prod elements} with a conjugation by $e^{-iPT}$.}  Most of the previous works assume that the observer can be minimally modeled as a clock \cite{Chandrasekaran:2022cip,Gomez:2022eui,Seo:2022pqj,Gomez:2023wrq,Gomez:2023upk,Jensen:2023yxy,Gomez:2023tkr}. 

In this work, we explore {the algebra of observables associated with certain `subsystems' of the full semiclassical theory. Following the change in the way the bulk observables are dressed, we see that these subsystems are described by a different type of von Neumann algebra than the one associated with the full semiclassical theory. We consider two cases to show this point and do so, first by studying bulk subregions, \emph{without having to add an observer by hand}, rather by considering the $T\Bar{T}$ deformation of the theory; and in the latter case, we study the operators accessible to an infalling observer from an asymptotic boundary.}  In the former, we adopt a well-known setting for holography in dS space (see \cite{Galante:2023uyf} for a review), {referred to} as interpolating geometries \cite{Anninos:2017hhn,Anninos:2018svg}. In the second case, we study the modifications in the algebra for general asymptotically AdS spacetimes. These two cases are examples where we see a transition in the algebra of observables from type II$_\infty$ to type II$_1$. However, they are a priori independent discussions.

The interpolating geometries are dilaton-gravity models that adopt near-AdS$_2$ space boundary conditions \cite{Maldacena:2019cbz}, {while the interior} is {a} near dS$_2$ space. {They avoid} a no-go theorem \cite{Freivogel:2005qh} forbidding any dS$_D$ region to reside in a causally accessible part of AdS$_D$ for $D>2$. We expect that a better understanding of the algebra of observables in these kinds of backgrounds will lead to new insights on their holographic dual theory \cite{Anninos:2020cwo}, and that of dS$_2$ JT gravity \cite{Susskind:2021esx,Susskind:2022dfz,Lin:2022nss,Lin:2022rbf,Susskind:2022bia,Rahman:2022jsf,Bhattacharjee:2022ave,Susskind:2023hnj}. 

JT gravity has been a productive test ground for studying of von Neumann algebras in gravity beyond the semiclassical regime \cite{Penington:2023dql,Kolchmeyer:2023gwa}, revealing the importance of different topologies in the description of the algebra of observables. However, the use of the centaur model in our work is not aimed at extending the discussion about the role of topologies, as above, but rather to emphasize the transition in the algebraic description of different subsystems of the semiclassical theory, in particular the $T\bar{T}$ deformations of the $2D$ flow geometries, which are expected to be associated with subregions in the bulks; and algebra of operators accessible to an observer given by a protocol according to \cite{Jafferis:2020ora}.   

\textbf{Structure}: In Sec. \ref{sec:set up} we start reviewing the semiclassical centaur geometry model, while in Sec. \ref{sec:algebra centaur} we study its crossed product algebra enlargement. Afterward, in Sec. \ref{sec:TTbar} we perform a $T\overline{T}$ deformation of the theory, where the gravitational constraints are imposed {by} the quasi-local Brown York energy, {resulting in modifications of the algebra}. Later, in Sec. \ref{sec:infalling obs} we study the experience of an infalling observer from the asymptotic boundary to the interior universe in the undeformed theory with the boundary theoretic protocol of \cite{deBoer:2022zps} {in} asymptotically AdS space of arbitrary dimensions. {In particular, the latter argument is also valid for the centaur geometries.} In the previous two cases, we focus on how the description of the observer changes the algebra from type II$_\infty$ to type II$_1$ and the conditions required for such modification. We conclude in Sec. \ref{sec:conclusions} with a brief summary of our main results and some future directions.

\section{Setting}\label{sec:set up}
{The first part of the manuscript is focused on} the 2-dimensional flow models \cite{Anninos:2017hhn,Anninos:2018svg,Anninos:2020cwo,Chapman:2021eyy,Anninos:2022hqo} which interpolate between an AdS$_2$ space and some internal space. In this section, we will briefly review the properties of these theories in Lorentzian signature, starting with the action
\begin{equation}\label{eq:action}
\begin{aligned}
    I=&I_0+\frac{1}{16\pi G_N}\int_{\mathcal{M}} \rmd^2x\sqrt{-{g}}\qty(\Phi R-V(\Phi))\\
    &+\frac{1}{8\pi G_N}\int_{\partial\mathcal{M}}\rmd x\sqrt{-{h}}\,K\Phi_b+I_m[g,\,\chi]~,
    \end{aligned}
\end{equation}
where $I_{0}$ represents a topological term, $\Phi$ is the dilaton field, $\Phi_b$ is the asymptotic boundary value of the dilaton, $\mathcal{M}$ is the spacetime manifold, $g_{\mu\nu}$ and $h_{\mu\nu}$ are the bulk and induced metric respectively; and $\chi$ represents the matter content of the theory, which is considered as a generic quantum field theory (QFT). The resulting equations of motion are given:
\begin{align}
    \nabla_\mu\nabla_\nu\Phi-g_{\mu\nu}\nabla^2\Phi-\frac{1}{2}g_{\mu\nu}V(\Phi)&=-8\pi G_N \expval{t_{\mu\nu}},\label{eq:EOM gmunu}\\
    R&=V'(\Phi)~,\label{eq:EOMunified}
\end{align}
where $\expval{t_{\mu\nu}}$ is the expectation value of the stress tensor for the matter fields, and the primes, $'$, indicate differentiation with respect to the argument of the function. In absence of such fields, $\epsilon^{\mu\nu}\partial_\nu\Phi$ is a Killing vector. Moreover, one can absorb the topological term of the action (\ref{eq:action}) in the definition of the dilaton $\Phi$ and expand the solution about $\Phi=\phi_0+\phi$. In the following, we work in the semiclassical limit $\phi_0\gg\phi$, since the dilaton represents the area of the transverse sphere of the higher dimensional near-Nariai black hole geometry.

\begin{figure}[t!]
    \centering
\includegraphics[width=0.4\textwidth]{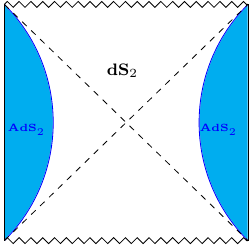}
    \caption{Lorentzian geometry of the dS$_2$ space, from the dimensional reduction of a near-Nariai black hole, patched to global AdS$_2$ space in cyan. Black dashed lines represent the dS horizon; and solid blue for the interpolating region. Figure based on \cite{Anninos:2022hqo}.
}
    \label{fig:Centaur}
\end{figure}

To describe the geometry, we also employ some particular dilaton potential term $V(\Phi)=2\Phi\tanh(\frac{\Phi}{\epsilon})$, where the $\epsilon\rightarrow0$ case represents a ``sharp" transition\footnote{This nomenclature, introduced in \cite{Anninos:2017hhn}, refers to the transition between AdS$_2$ and dS$_2$ spaces to be localized $\Phi=0$, while the case with $\epsilon$ finite takes place over a finite interval for $\Phi$, which is referred to as a ``smooth" transition.} between AdS space and the interior geometry, which can be AdS$_2$ or dS$_2$ space depending on the sign of the renormalized dilaton. For concreteness, we focus on the case where $\Phi_b>0$ to obtain a transition between spacetimes of opposite sign curvature. In that case, the potential becomes,
\begin{equation}\label{eq:V AdS}
    V_{\rm cent}(\Phi)=2\eta{\Phi}+\tilde{\phi}~;
\end{equation}
where
\begin{equation}
    \eta=\begin{cases}
       +1&\text{AdS}_2~,\\
       -1&\text{dS}_2~.
    \end{cases}
\end{equation}
An illustration of this geometry is shown in Fig. \ref{fig:Centaur}.

This construction is a double-sided geometry, i.e. a particle on each of the boundaries is required to describe the bulk geometry. It becomes convenient to introduce the conformal metric 
\begin{equation}\label{eq:coord rho tau}
\rmd s^2=\rme^{2\omega(\rho,\,\tau)}(-\rmd\tau^2+\rmd\rho^2)    
\end{equation}
with $\omega(\rho,\,\tau)$ the conformal factor. Explicitly: 
\begin{equation}\label{eq:e 2 omega}
    \rme^{2\omega(\rho,\,\tau)}=\begin{cases}
        \sec^2\rho~,&0\leq\rho\leq\frac{\pi}{2}~,\\
        \sech^2\rho~,&-\infty<\rho\leq0~,
    \end{cases}
\end{equation}
where the AdS$_2$ boundary corresponds to $\rho=\pi/2$.

Let us consider a curve of the form $\mathcal{C}=\qty{\tau(u),\,\rho(u)}$ to parametrize the embedding of one of the boundaries (say R). We impose Dirichlet boundary conditions in $\mathcal{C}$, by scaling $\Phi_b(u)$ and $h(u)$ with $\Lambda\gg1$ as $\qty[\sqrt{h},\,\Phi_b]\rightarrow \Lambda\qty[1,\,\Phi_r]$. The resulting on-shell action of (\ref{eq:action}):
    \begin{equation}\label{eq:bdyI}
      I_{\rm on} = \frac{1}{8\pi G_N}\int\rmd u\,\Phi_r(u)\qty(\Lambda^2+\frac{1}{2}(\tau'(u))^2+\qty{\tau(u),\,u})~,
    \end{equation}
where 
\begin{equation}
    \qty{\tau(u),\,u}=\frac{\tau'''(u)}{\tau'(u)}-\frac{3}{2}\qty(\frac{\tau''(u)}{\tau'(u)})^2
\end{equation}
is the Schwarzian derivative. (\ref{eq:bdyI}) follows from the original derivation in \cite{Anninos:2018svg} of the Schwarzian theory for the centaur geometry in Euclidean signature after the appropriate Wick rotation. 

Note that the term $\Lambda^2$ can be eliminated with standard holographic renormalization \cite{deHaro:2000vlm,Bianchi:2001kw}. After eliminating this term, we can perform the change of variable $\sigma(u)=\tan\frac{\tau(u)}{2}$ to modify (\ref{eq:bdyI}) into
\begin{equation}\label{eq:bdyI2}
      I_{\rm on} = \frac{1}{8\pi G_N}\int\rmd u\,\Phi_r(u)\qty{\sigma(u),\,u}~.
    \end{equation}
Since $\tau(u)$ in Lorentzian signature is non-compact, this means that the time isometries (\ref{eq:bdyI2}) correspond to SL$(2,~\mathbf{R})$ symmetries of the theory,
\begin{equation}
    \sigma(u)\rightarrow\frac{a_1\sigma(u)+a_2}{b_1\sigma(u)+b_2}~,
\end{equation}
where $a_1,~a_2,~b_1$ and $b_2\in\mathbf{R}$ and $a_1a_2-b_1b_2=1$. The symmetry group then coincides with that one of global (nearly) AdS$_2$ space \cite{Maldacena:2016upp}.

The one-sided Hamiltonian, $H_{\rm cent}$, corresponding to the boundary action (\ref{eq:bdyI}) can be deduced as \cite{Woodard:2015zca,Gross:2019ach}
\begin{equation}
\begin{aligned}
    H_{\rm cent}&=\frac{\phi_r(u)}{8\pi G_N}\qty(\frac{\tau'(u)^2}{2}-\frac{\tau'''(u)}{\tau'(u)}+\frac{3}{2}\qty(\frac{\tau''(u)}{\tau'(u)})^2)~.\label{eq:Mod Hamiltonian algebraic}
\end{aligned}
\end{equation}
The centaur geometry is a maximally extended asymptotically AdS spacetime. We can define a TFD state for the boundary theories on the left and right boundaries of Fig. \ref{fig:Centaur}, where the one-sided Hamiltonians on the $L$ and $R$ act with a suitable boost. The algebra of the $L$ or $R$ boundary theory without matter consists of bounded functions of $H_{\rm cent,\,R}$ or $H_{\rm cent,\,L}$ respectively. In this way, one can define the modular Hamiltonian $H_{\rm cent,\,R} -H_{\rm cent,\,L}$ that preserves the time translation in the TFD construction from the same arguments as \cite{Witten:2021unn,Chandrasekaran:2022cip}.

It also suffers from a similar factorisation puzzle as in JT gravity \cite{Harlow:2018tqv,Penington:2023dql,Kolchmeyer:2023gwa}, namely that the algebras $\mathcal{A}_L$ and $\mathcal{A}_R$ commute with each other, since they share the same generator, $H_{\rm JT,\,L}=H_{\rm JT,\,R}$. In the case of the centaur geometry, the SL$(2,~\mathbf{R})$ invariance corresponds to the modular Hamiltonian $H_{\rm mod}=H_{\rm cent,\,L}-H_{\rm cent,\,R}$, and the Hamiltonian constraint on physical states $\ket{\psi}$ which are invariant under this symmetry reads $H_{\rm mod}\ket{\psi}=0$, which expresses that $H_{\rm cent,\,L}=H_{\rm cent,\,R}$ for physical states. This issue is no longer present once that matter is introduced, as follows. We define local matter operators, $\chi$, with the appropriate canonical quantization relations respecting SL$(2,~\mathbf{R})$ gauge invariance. A careful treatment of the algebra of observables for SL$(2,~\mathbf{R})$ coinvariant states has been presented in \cite{Penington:2023dql} in the context of AdS$_2$ JT gravity. Note that even though the procedure for defining the algebra of operators in the same as the JT gravity case, the bulk fields are now propagating in a different background and the mode expansion for these fields will be different from the one on the JT background.

The smearing over $u$ allows us to define bounded operators $\mathcal{B}(\chi)$. We can then express the time translation generators along the $L/R$ boundary as
\begin{equation}\label{eq:One sided H}
    H_{\rm L/R}=H_{\rm cent,\, L/R}+H_{\rm matter,\, L/R}~,
\end{equation}
where the generator of SL$(2,~\mathbf{R})$ transformations corresponds to the modular Hamiltonian,
\begin{equation}
    H=H_R-H_L~.\label{eq:non trivial generator}
\end{equation}
Once we add matter to the theory, we can employ a generalized free-field approximation for constructing the total Hilbert space $\mathcal{H}_{\rm tot}$,
\begin{equation}
    \mathcal{H}_{\rm tot}=\mathcal{H}_{\rm matt}\otimes \mathcal{H}_{\rm grav}~,\label{eq:Hilbert space total}
\end{equation}
where the operators quantizing the metric and the dilaton $\qty{h_{\mu\nu}^{\rm grav},\,\phi}$ can be used to construct the states in $\mathcal{H}_{\rm grav}$; meanwhile, $\mathcal{H}_{\rm matter}$ can be constructed from strings of Fourier modes $\qty{a,\,a^\dagger}$, i.e. any matter field $\chi$ can adopt a decomposition
\begin{equation}\label{eq:quantization of chi}
    \chi(\tau(u))=\int \rmd\omega\,(f_\omega(\tau(u))a_\omega+f^*_\omega(\tau(u))a^\dagger_\omega)~.
\end{equation}
We notice that since we consider a generic quantum field theory on a curved spacetime background, the algebra of operators consisting of combinations of $a_{\omega}$ and $a^\dagger_{\omega'}$ will be described by a type III$_1$ von Neumann algebra.

As a remark, although the centaur and JT gravity share the same symmetry algebra (SL$(2,~\mathbf{R})$), the algebra of operators
is different due to the presence of the dS bulk interior. This is contained in the coefficients of $f_\omega(\tau(u))$, $f_\omega^*(\tau(u))$ as they are subject to appropriate boundary conditions. We refer the reader to the Klein-Gordon equation in the sharp centaur geometry in \cite{Anninos:2017hhn}. There are different types of modes, depending on the frequency $\omega$. An important difference is that they tend to be less efficient at thermalizing with respect to the pure AdS$_2$ case. This means that the difference in the bulk geometry is encoded in the operators $\chi$ fields, which are members in the algebra $\mathcal{A}$.

We will explore the consequences of the crossed product in the following sections.

\section{Algebra for the centaur geometry}\label{sec:algebra centaur}
The full boundary algebra for a given side, such as $R$, is generated by $H_R$ and $\chi$. This determines the type III$_1$ algebra of operators by constructing finite strings of the modes $\qty{a,\,a^\dagger}$ and bounded functions of $H_R$. {To incorporate gravity in our discussion, we treat time translations as gauge redundancies in the context of AdS/CFT correspondence.} Let us denote the Hilbert space of the coinvariant (i.e. defined by equivalence classes) state as $\hat{\mathcal{H}}$ by all SL$(2,~\mathbf{R})$ invariant renormalizable states constructed from the operators (\ref{eq:Hilbert space total}) and their Hilbert space completion. Details about the construction of this particular Hilbert space can be found in \cite{Penington:2023dql}.

Let $\mathcal{A}_R$ be the von Neumann algebra that consists of the set of operators $\qty{\hat{\mathcal{O}}_R}$ in $R$ which time evolve non-trivially along the asymptotic boundary by the modular flow (\ref{eq:non trivial generator}) according to (\ref{eq:U ops}) with $U=\rme^{\rmi H \tau}$, where $H$ is the difference between the left and right Hamiltonians of the centaur geometry. We start from a thermofield-double state, $\ket{\psi_{\rm TFD}}$, of the boundary theories in Fig. \ref{fig:Centaur}, which is a cyclic and separating vacuum state that obeys the constraint equation
\begin{equation}\label{eq:One sided H constraint}
    H\ket{\psi_{\rm TFD}}=0~.
\end{equation}
Then, we can generate cross-product algebra following (\ref{eq:crossed prod elements}) with $T=H$, i.e. the modular time translation generator of the cross-product algebra; and $X=H_L$\footnote{Notice however, this identification is purely formal since $X\in\mathcal{A}'$, while $H_L\notin\mathcal{A}'$ for type III theories.}. {Given the SL$(2,~\mathbf{R})$ invariance, the automorphism group is $\mathbb{R}$ as in (\ref{eq:crossed prod}).} Consider now
\begin{equation}
\hat{a}_R\in \hat{\mathcal{A}}_{\rm R}\,.
\end{equation}
Since $H\ket{\psi_{\rm TFD}}=0$, (\ref{eq:One sided H constraint}) can be employed to evaluate the expectation value of a generic element $\hat{a}\in \hat{\mathcal{A}}_{\rm R}$ \cite{Witten:2021unn},
\begin{equation}
    \Tr \hat{a}=\beta_{\rm TFD}\int_{X_{\rm min}}^{X_{\rm max}}\,\rmd X\,\rme^{\beta_{\rm TFD}X}\bra{\psi_{\rm TFD}}a(X)\ket{\psi_{\rm TFD}}\,.\label{eq:Trace ahat}
\end{equation}
We have introduced the integration limits $X_{\rm min}$ and $X_{\rm max}$ to indicate constraints in the one-sided modular Hamiltonian. In the present case, given that time isometries with respect to the asymptotic boundary correspond to the SL$(2,~\mathbf{R})$ group, the allowed range becomes $X\in(-\infty,\infty)$. 

The definition (\ref{eq:Trace ahat}) obeys the properties
\begin{equation}\label{eq:traces}
\begin{aligned}
    \Tr\hat{a}\hat{b}=\Tr \hat{b}\hat{a}&\qquad \hat{a},~\hat{b}\in\hat{\mathcal{A}}~,\\
    \Tr \hat{a}^\dagger\hat{a}>0&\qquad\forall\, \hat{a}\neq 0~.
\end{aligned}
\end{equation}
As mentioned in the introduction, the crossed product will result in a type II algebra. Moreover, since the trace of the identity matrix $\Tr \mathbb{1}\rightarrow\infty$ then the algebra is type II$_\infty$. The trace must be finite for a dense set of operators in the algebra.

\section{\texorpdfstring{$T\overline{T}$}{} deformed theory}\label{sec:TTbar}
Our goal in this section is to try to address the algebra of observables in a bounded subregion \cite{Bahiru:2022mwh,Leutheusser:2022bgi,Jensen:2023yxy,AliAhmad:2023etg} for the centaur geometry by implementing a $T\overline{T}$ deformation \cite{McGough:2016lol,Kraus:2018xrn,Donnelly:2018bef,Hartman:2018tkw,Gross:2019ach,Gross:2019uxi}. 

First, the symmetry group is preserved under $T\overline{T}$ deformations \cite{Gross:2019ach}, so we restrict the discussion about the algebra of operators using the Hilbert space of coinvariants of SL$(2,~\mathbf{R})$, as in Sec. \ref{sec:algebra centaur}. 

We follow the conventions \cite{Gross:2019ach} to express $T\overline{T}$ deformations parametrized by $\lambda\in\mathbb{R}$ as
\begin{equation}
    \dv{I}{\lambda}=\pi\int \rmd^2x\,\sqrt{-g}\qty( T^{ij}T_{ij}-(T^i_i)^2)~,\label{eq:general defor}
\end{equation}
where $T_{ij}$ is the Brown-York quasilocal stress tensor \cite{Brown:1992br,Balasubramanian:1999re} along a boundary surface $r=\frac{1}{\sqrt{\alpha{\lambda}}}$, with $\alpha\equiv 1/(2\pi G_N)$, in static patch coordinates
\begin{equation}
    \rmd s^2=-N(r){\rmd\tau^2}+\frac{\rmd r^2}{N(r)}~,\quad \Phi=\Phi(r)~,
\end{equation}
in the absence of matter; while equation (\ref{eq:general defor}) and the relation between the cutoff and $\lambda$ is modified in presence of matter \cite{McGough:2016lol,Kraus:2018xrn}.
Alternatively, the deformation can be interpreted as the result of introducing mixed boundary conditions for the undeformed theory \cite{Guica:2019nzm}. In the former interpretation, the time translation generator along the left or right-sided cutoff surface is given by the quasi-local Brown-York Hamiltonian, $H_{T\overline{T}}$. For a general dilaton-gravity theory with matter of the form (\ref{eq:action}) under the $T\overline{T}$ deformation (\ref{eq:general defor}), the quasi-local Brown-York Hamiltonian obeys the relation \cite{Gross:2019ach}
\begin{equation}\label{eq:spectrum energies}
    \pdv{H_{T\overline{T}}}{\lambda}=\frac{H_{T\overline{T}}^2-\frac{1}{16\lambda^2}\qty(1+\frac{\sqrt{\alpha\lambda}}{2\Phi_r}V\qty(\frac{\Phi_r}{\sqrt{\alpha\lambda}}))-\frac{t_r^r}{4\alpha^{1/2}\lambda^{3/2}}}{1/2-2\lambda H_{T\overline{T}}}~,
\end{equation}
where ${t}_r^r$ is the radial-radial component of the bulk matter stress tensor at the cutoff surface, and $H_{T\overline{T}}(\lambda=0)=H_{\rm cent}$.

The precise deformation of the energy spectrum will depend on the details of the matter stress tensor, the dilaton potential, and the location where the cutoff is performed. For example, consider the centaur theory with the potential (\ref{eq:V AdS}) and $t_r^r=$const in (\ref{eq:spectrum energies}),
\begin{equation}\label{eq:TTbar spec}
    H_{\rm cent}^{\lambda}\approx \frac{1}{4\lambda}\qty(1-\sqrt{\eta-8\sqrt{\frac{\lambda}{\alpha}}t_r^r-8\lambda H_{\rm cent}})~,
\end{equation}
with $H_{\rm cent}$ is the modular Hamiltonian of the undeformed theory (\ref{eq:Mod Hamiltonian algebraic}). 

Given the SL$(2,~\mathbf{R})$ time isometry in the modular Hamiltonian, the spectrum of $H_{\rm cent}\in(-\infty,~\infty)$. However, there is the presence of a square root in (\ref{eq:TTbar spec})\footnote{On the other hand, one can also find a complex energy spectrum in (\ref{eq:TTbar spec}) for $H_{\rm cent}$ large enough. This situation is quite generic for $T\overline{T}$ deformed theories \cite{Gross:2019ach}. These energies are commonly discarded by imposing unitarily in the resulting theory.}, which indicates that $X$ will be bounded from above. This means that the definition of the trace (\ref{eq:Trace ahat}) will be finite under the $T\overline{T}$ deformation. Moreover, notice that (\ref{eq:TTbar spec}) would produce the spectrum of a $T\overline{T}+\Lambda_2$ deformation \cite{Gorbenko:2018oov,Lewkowycz:2019xse,Coleman:2021nor} directly in the interpolating geometry.

Notice that our result is consistent with the prediction of \cite{Jensen:2023yxy} for closed universes. {In this derivation, we worked under the assumption that} the stress tensor is a bounded function; it would be interesting to study if this restriction can be relaxed.

\section{The experience of an infalling observer}\label{sec:infalling obs}
We employ the protocol of \cite{Jafferis:2020ora,Gao:2021tzr,deBoer:2022zps} that describes the experience of an infalling observer, modeled as a probe black hole, from the boundary of a generic asymptotically AdS$_{d+1}$ spacetime{, including the centaur geometry}. We prepare a microcanonical TFD configuration dual to a black hole geometry and a copy of it, which we refer to as the reference system, with energy eigenstates $\ket{E_n}_{\rm sys}$ and $\ket{\overline{E}_n}_{\rm ref}$ respectively. We employ a conformal transformation $\rme^{\rmi P\rho}$ to shift the black hole into the asymptotic boundary, with $P$ the momentum operator, and $\rho\gg1$ a parameter controlling the shift. Let $\ket{\psi}$ denote the CFT state dual to a semiclassical asymptotically AdS space with a probe black hole. Defining the state \cite{deBoer:2022zps}
\begin{equation}
\begin{aligned}
    \ket{\psi}=Z^{-1/2}\sum_n&f(E_n|E_0,\,\sigma)\qty[V_{\rm sys}\rme^{-\delta\ell^2P^2-\rmi P\rho}\ket{E_n}_{\text{sys}}]\ket{\overline{E}_n}_{\text{ref}}~,
\end{aligned}
\end{equation}
where $V_{\rm sys}$ is some arbitrary operation in the interior geometry; $f(E_n|E_0)$ is an appropriate enveloping function for $E_n$ to be summed over a microcanonical window of width $\sigma$ around $E_0$; $\delta\ell$ is the wavepacket localization; {and $Z$ the microcanonical partition function}.

The set of normalizable states $\ket{\psi}=\qty{\ket{\psi_{\rm eq}}}$ are called local equilibrium operators, which by definition obey KMS conditions for the two-point functions of the set of operators available to the atmosphere around the observer, denoted by $O=\qty{\phi_{\rm atm}}$:
\begin{equation}
    \bra{\psi_{\rm eq}}O_{1}^\dagger\exp[-2\pi K_{\rho_{\rm eq}}]O_2\ket{\psi_{\rm eq}}=\bra{\psi_{\rm eq}}O_2O_1^\dagger\ket{\psi_{\rm eq}}~,\label{eq:KMS}
\end{equation}
where $K_{\rho_{\rm eq}}$ is the modular Hamiltonian,
\begin{equation}\label{eq:K eq}
    K_{\rho_{\rm eq}}=-\frac{1}{2\pi}\log\qty[\rho_{\rm eq}]~,
\end{equation}
with $\rho_{\rm eq}=\ket{\psi_{\rm eq}}\bra{\psi_{\rm eq}}$. Then, the generator of Schwarzschild time translation for the proper time of these states is given by tracing out the reference system $K^{\rm sys}_{\rho_{\rm eq}}=\hat{\Tr}_{\rm ref}K_{\rho_{\rm eq}}$.

On the other hand, since the reference system is entangled by construction with the infalling observer; it is then natural to employ $K^{\rm ref}_{\rho_{\rm eq}}=\hat{\Tr}_{\rm sys}K_{\rho_{\rm eq}}$ as the time automorphism generator for the infalling observer. By making this identification, we employ $T=K_{\rho_{\rm eq}}$ as the generator of time automorphism of $\hat{\mathcal{A}}$, and $X=K^{\rm ref}_{\rho_{\rm eq}}$. This allows us to define the traces of the crossed product algebra in (\ref{eq:Trace ahat}). A specific example where reduced density matrices in the microcanonical ensemble have been explicitly constructed in the infalling bulk observer protocol can be found in \cite{Gao:2021tzr} where the boundary theory is taken as the SYK model.

{We focus on the case where the infalling probe black hole does not encounter bulk matter fields along its worldline. Its experienced Hilbert space is then always described by equilibrium states.} In such a case, we must also account for the constraint that the reference system energy is bounded from below. This comes from the construction of the generator (\ref{eq:K eq}). Given that $\ket{\psi_{\rm eq}}$ obey the KMS relation (\ref{eq:KMS}) for all elements in the algebra $\mathcal{A}$, they are normalizable states. It is then clear that the range of integration $[X_{\rm min},\,X_{\rm max}]$ in (\ref{eq:Trace ahat}) is a bounded interval, and as such the trace is finite $\forall\hat{a}\in\hat{\mathcal{A}}$, i.e. the von Neumann algebra is thus type II$_1$. 

{In general,} the nature of the algebra will be determined by the states in $\hat{\mathcal{H}}$. The presence of matter in the background geometry {can introduce} non-equilibrium states. Such states {are in principle} non-normalizable, {leading to ill-defined traces} for some elements in $\mathcal{A}$. {Thus,} the experience of infalling observer {can still be described} with type II$_\infty$ algebras {generically; although, considering} symmetries of the system might result in a type II$_1$ description.

Notice that the particular use of two-dimensional gravity was not employed in the arguments, in fact, we have not used any input from the background geometry in our arguments. The only requirement to get the type II$_{1}$ algebra is that the state be an equilibrium state. Therefore, the construction works for general asymptotically AdS$_{d+1}$ spacetimes without matter. The transition occurs as soon we exchange the ADM Hamiltonian for the reference system $K^{\rm ref}_{\rho_{\rm eq}}$ by dressing the bulk operators with respect to the probe black hole.

\section{Conclusions and outlook}\label{sec:conclusions}
In this work, we have uncovered the transition in the type II algebra of observables by considering (i) a $T\overline{T}$ deformation {of the centaur geometries}, and (ii) the experience of an infalling observer from the boundary to the interior geometry {of asymptotically AdS spacetimes}. {In both cases,} the transition {to a type II$_1$ algebra} allows us to construct a maximally mixed state and a notion of generalized entropies. Meanwhile, in the latter case, we have shown that {if the infalling observer from the asymptotic boundary does not cross bulk matter fields}, the transition of the algebra type II$_\infty$ to type II$_1$ does not depend on the interior geometry, as long as the protocol \cite{deBoer:2022zps} can be employed.

Let us proceed by pointing out some future directions.

First, as we have indicated, after the crossed product enlargement of the algebra (\ref{eq:crossed prod}), the definition of traces in (\ref{eq:traces}) allows us to define reduced-density matrices and rigorous notions of generalized entropies. Interesting progress towards formulating the Page curve in the language of von Neumann algebras was initiated in \cite{Verlinde:2022xkw,Gomez:2023tkr}. Perhaps such notions can establish the island formula on solid grounds for de Sitter space, which was pioneered by \cite{Svesko:2022txo}. However, it has been argued that the appearance of islands close to the cosmological horizon violates entanglement wedge nesting \cite{Shaghoulian:2021cef} unless the large backreaction is induced \cite{Aalsma:2021bit,Aalsma:2022swk}. We hope the algebraic techniques can bring a better understanding of these features.

Second, one might generalize the lessons from our study on $T\overline{T}$ deformation of the centaur more broadly to dilaton-gravity theories of the form (\ref{eq:action}) where the spectrum of quasi-local energies at the cutoff surface in (\ref{eq:spectrum energies}) remains bounded. Perhaps, the simplest explicit generalizations would involve the AdS$_2$ interpolating geometries with a different cosmological constant; the $\gamma$-centaur \cite{Anninos:2018svg}, and the double interpolating geometry \cite{Anninos:2022hqo}; where the change in the algebra is also suggested by \cite{Jensen:2023yxy}.

Third, although the centaur geometries provide a natural background to study de Sitter space holography and a rich algebraic structure; these theories are known to be thermodynamically unstable \cite{Anninos:2017hhn}, which motivated the construction of the double interpolating geometries in \cite{Anninos:2022hqo}. It would be interesting to study the thermodynamic properties of the $T\overline{T}$ deformed centaur geometry, as they have not received much attention since the original work of \cite{Gross:2019ach}. However, the energy spectrum is generically complex under these deformations. Although one can restrict the energy eigenstates to describe a unitary theory, a new perspective arises with Cauchy slice holography \cite{Araujo-Regado:2022gvw,Araujo-Regado:2022jpj} where the notion of complex stress tensor plays a crucial role, which would be interesting to study explicitly for the centaur geometry, and whether the restriction on the finiteness of the radial-radial component of the matter stress tensor $t^r_r$ can be lifted and still recover the transition in the algebra.

Fourth, we can think of the infalling observer as a little diary falling into a black hole. Certain protocols can be used to recover the information after the scrambling time \cite{Hayden:2007cs, Yoshida:2017non}. In the context of the Page curve, the information encoded in the island can be recovered by applying explicit teleportation protocols \cite{Almheiri:2019yqk, Aguilar-Gutierrez:2023ymx}, see upcoming work in this area by \cite{toappear1, vdHV}. It would be interesting to understand how information recovery works in algebraic language. These ideas can shed some light on understanding the microscopic origin of the island formula.

Moreover, as we have emphasized, our result for the infalling observer does not depend on the specific interior geometry. It would be interesting to understand the relation of background independence with other approaches such as \cite{Witten:2023xze}. However, the notion of equilibrium states that were employed to define the reduced density matrices with respect to the observer in the boundary theory protocol of \cite{deBoer:2022zps} relies on the same original assumptions, in particular, that the equilibrium states need to minimize a notion of circuit complexity in the boundary theory, which has not been developed explicitly so far. We hope that the algebraic techniques uncovered in this work can catalyze progress on rigorously defining complexity proposals from the boundary perspective and the respective bulk realization, {initiated in \cite{Erdmenger:2022lov}}. In that case, the centaur geometry could be a productive test ground for the different proposals for holographic complexity \cite{Chapman:2021eyy} in stretched horizon holography \cite{Jorstad:2022mls,Anegawa:2023wrk,Baiguera:2023tpt,Anegawa:2023dad,Aguilar-Gutierrez:2023zqm,Aguilar-Gutierrez:2023tic}, and possibly to incorporate quantum corrections in such proposals, as recently studied in \cite{Carrasco:2023fcj}.

Finally, it would be interesting to incorporate non-equilibrium states in the protocol of the infalling observer to study modification in the algebra of observables for the probe black hole. The probe will absorb particles along its worldline. Then, the crossed product algebra could remain type II$_\infty$ instead of II$_1$, as traces might include non-normalizable states. Regardless of that, the evolution of the atmosphere operators in the algebra will be determined by scrambling modes of the modular Hamiltonian \cite{Jafferis:2020ora}. Moreover, these modes produce null shifts along the horizon of the background black hole \cite{deBoer:2022zps}. This could allow for a wormhole teleportation protocol for the {probe} black hole, {seen as a diary}. It might be worth studying the algebraic structure of such protocol explicitly with an SYK model dual to a near AdS$_2$ space, as first proposed in \cite{Gao:2021tzr}.

\section*{Acknowledgments}
We would like to thank Shadi Ali Ahmad, Dio Anninos, Damián Galante, Stefan Hollands, Ro Jefferson, Andrew Rolph, Sirui Shuai, Andrew Svesko, Eleanor Harris, and Yixu Wang for useful discussions on centaur spacetimes and von Neumann algebras, {and specially Manus Visser for early collaboration}. SEAG thanks the IFT-UAM/CSIC, the University of Amsterdam, the Delta Institute for Theoretical Physics, and the International Centre for Theoretical Physics for their hospitality and financial support during several phases of the project, and the Research Foundation - Flanders (FWO) for also providing mobility support. EB also wants to thank the CERN-TH for their hospitality during the preparation of this paper. The work of SEAG is partially supported by the KU Leuven C1 grant ZKD1118 C16/16/005. The work of EB is partially supported by the Eramsus+ Trainee-ship programme and the INFN Iniziativa Specifica String Theory
and Fundamental Interactions. RE is supported by the Dushi Zhuanxiang Fellowship and acknowledges a Shuimu Scholarship as part of the “Shuimu Tsinghua Scholar” Program.

\bibliographystyle{JHEP}
\bibliography{references.bib} 

\end{document}